\documentclass{article}
\usepackage{amsmath,amssymb,amsthm,amsfonts}

\usepackage{bbm,dsfont}
\usepackage{graphicx,epsfig,color}
\usepackage[margin=3cm]{geometry}

\renewcommand{\thefootnote}{\arabic{footnote}}
\numberwithin{equation}{section}

\newcommand{\id}{{\boldsymbol{\mathbbm{1}}}}
\DeclareMathOperator{\tr}{tr}
\DeclareMathOperator{\sym}{sym}
\DeclareMathOperator{\dev}{dev}
\DeclareMathOperator{\skewp}{skew}

\DeclareMathOperator{\Curl}{Curl}
\DeclareMathOperator{\grad}{grad}
\DeclareMathOperator{\divv}{div}
\DeclareMathOperator{\Divv}{Div}
\DeclareMathOperator{\polar}{polar}

\usepackage{tikz}

\begin{document}

\title{Geometrically nonlinear Cosserat elasticity in the plane: applications to chirality}
\author{\renewcommand{\thefootnote}{\arabic{footnote}}  
  Sebastian Bahamonde\footnotemark[1] \ and
  Christian G. B\"ohmer\footnotemark[2] \ and
  Patrizio Neff\footnotemark[3]}
\date{\today}

\footnotetext[1]{Sebastian Bahamonde, Department of Mathematics, University College London, Gower Street, London, WC1E 6BT, UK, email: sebastian.beltran.14@ucl.ac.uk}

\footnotetext[2]{Corresponding author: Christian G. B\"ohmer, Department of Mathematics, University College London, Gower Street, London, WC1E 6BT, UK, email: c.boehmer@ucl.ac.uk}

\footnotetext[3]{Patrizio Neff, Fakult\"at f\"ur Mathematik, Universit\"at Duisburg-Essen, Thea-Leymann-Stra\ss e 9, 45127 Essen, Germany, email: patrizio.neff@uni-due.de} 

\maketitle

\begin{abstract}
  Modelling two-dimensional chiral materials is a challenging problem in continuum mechanics because three-dimensional theories reduced to isotropic two-dimensional problems become non-chiral. Various approaches have been suggested to overcome this problem. We propose a new approach to this problem by formulating an intrinsically  two-dimensional model which does not require references to a higher dimensional one. We are able to model planar chiral materials starting from a geometrically non-linear Cosserat type elasticity theory. Our results are in agreement with previously derived equations of motion but can contain additional terms due to our non-linear approach. Plane wave solutions are briefly discussed within this model.
\end{abstract}

\mbox{}\\ 
\mbox{}


\mbox{}


\mbox{}

\section{Introduction}

Classical elasticity theory assumes structureless material points. These points are not allowed to possess an additional so-called microstructure which could take into account properties like orientation or volume of the material points. It is possible to extend the theory of classical elasticity to take into account this additional structure, this is known as the Cosserat~\cite{Cosserat09} model. In the Cosserat continuum, material points can, for instance, experience rotations without stretches. Therefore, in addition to the standard deformation field $\varphi$ there is an independent rotation field $\overline{R}$, which means that $\overline{R}$ is an orthogonal matrix. Many models in continuum mechanics were motivated by this idea which has resulted in many interesting research lines, sometime with varying names~\cite{Ericksen:1957,Toupin:1962,Ericksen:1962a,Ericksen:1962b,Mindlin1964,Toupin:1964,Eringen:1964a,Eringen:1964b,Green:1964,Ericksen:1967,Schaefer:1967,Eringen:1999}

The three-dimensional static nonlinear Cosserat model has seen a tremendous increase of interest in recent years~\cite{Neff_Cosserat_plasticity05,Neff_Muench_simple_shear09,Neff_Muench_transverse_cosserat08,Neff_Biot07,2016arXiv160306633F,2016arXiv160609085B,2017arXiv170108150F}. This is connected to its possibility to model uncommon effects like for instance lattice rotations. Working with the manifold $\mathrm{SO}(3)$ of proper rotations required, however, many new tools from the mathematical and implementational side. Expositions regarding the mathematical treatment can be found in~\cite{Neff_micromorphic_rse_05,Neff_curl06,Neff_Gamm04,Neff_Osterbrink_Cosserat15,2015arXiv150408003L}, similarly for computational results in~\cite{Fischle_Neff_optimal16_2,Fischle_Neff_optimal16_1}. The Cosserat approach is maybe best known for its ability to model thin shell structures. Here, the additional orthogonal frame provided by the Cosserat theory fits well into the theory of deformable surfaces~\cite{Neff_plate04_cmt,Neff_plate05_poly,Neff_Hong_Reissner08,Birsan_Neff_JElast2013,Birsan_Neff_MMS_2013,Sander_shell16}. Not much is known, regrettably, in the case of general nonlinear dynamics. Our contribution~\cite{Boehmer2016158} sheds some light in this direction. 

One of the major unresolved issues of the theory is the precise matter in which elastic deformations (macroscopic) and Cosserat microrotations are coupled. If we consider a quadratic ansatz in the stretch tensor, it usually comes down to writing a coupling term of the form $\mu_c \|\skewp(\overline{U}-\id)\|^2$ in which $\mu_c \geq 0$ is known as the Cosserat couple modulus. The effect of this coupling term is clear by increasing $\mu_c \rightarrow \infty$. Then the Cosserat rotations become the continuum rotations $R=\polar(F)$. A similar effect can be obtained by instead considering the coupling $\mu_c \|\overline{R}^T \polar(F)-\id\|^2$. Both terms induce the same linear response. It is important to note that the geometrically nonlinear Cosserat model can be used also with $\mu_c \equiv 0$, a possibility which is meaningless in the linear Cosserat model. In the case $\mu_c=0$ one has a Cosserat model with symmetric stresses, which stretches may be non-symmetric.

Cosserat elasticity is generally formulated as a three-dimensional continuum mechanics theory and planar problems are usually considered by restricting either displacements or microrotations to the plane. One sometimes speaks of the first planar problem when $\mathbf{u}=(u_1,u_2,0)$ and rotations are constrained to be about the $z$-direction. The second planar problem deals with the opposite situation where $\mathbf{u}=(0,0,u_3)$ while the rotations are constrained about the $x$-axis and $y$-axis, see for instance~\cite{Jasiuk95,Joumaa20112896}.

Chirality\footnote{``I call any geometrical figure, or group of points, chiral, and say that it has chirality if its image in a plane mirror, ideally realised, cannot be brought to coincide with itself.'' (Lord Kelvin 1894)} or handedness is a common feature in various fields of science. It refers to the possibility of an object or system to be distinguishable from its mirror image. Many molecules in chemistry are chiral, most often due to the presence of an asymmetric carbon atom. Chiral material have been of interest in continuum mechanics since the 1980s. When one is interested in studying chiral materials~\cite{Lakes19821161,Lakes20011579}, it turns out that a three-dimensional theory when reduced to an isotropic two-dimensional problem becomes non-chiral~\cite{Lakes20011579,Rosi2016120,Liu20121907}.

Typically, a chiral term in a three-dimensional elasticity model would be of the form
\begin{align}
  (\overline{R}^T F)_{ij}\, \mathbb{C}_{ijmn}\,
  (\overline{R}^T\! \Curl \overline{R})_{mn} \,,
\end{align}
where $C_{ijmn}$ is a material tensor, see also Appendix~\ref{app:chiral2}. If one considers an isotropic material tensor of the form $c_1 \delta_{ij}\delta_{mn} + c_2 \delta_{im}\delta_{jn} + c_3 \delta_{in}\delta_{jm}$, then the above term yields three contributions. It turns out that these three terms identically vanish when the first Cosserat planar problem is considered\footnote{It might be possible to construct chiral terms using non-linear functionals beyond the usual quadratic terms which yield a non-trivial planar theory.}. For instance, based on representation theorems~\cite{Cheverton1981}, a total of 20 invariants were discussed, five of which are chiral according to our formulation in Appendix~\ref{app:chiral2}. However, a direct calculation verifies that all chiral terms vanish when the deformation gradient is confined to the plane, see Eq.~(\ref{p1}).

Hence, studying two-dimensional chiral materials requires a new approach. One such approach is the use of strain gradient theories, see~\cite{Rosi2016120}, for instance. Another approach goes back to~\cite{Liu20121907} where chirality was introduced in the two-dimensional setting by revisiting isotropic 4th order tensors and identifying an extra piece in the constitutive relation. It was subsequently shown that this additional part of the elastic tensor can indeed give a meaningful model with chirality. The model considered in~\cite{Liu20121907} is based on linear Cosserat elasticity, see also~\cite{10.1117/12.915038,sv-jmesv-jme.2016.3799}.

In the present paper, we approach this problem from a very different point of view. We begin by carefully studying the basic formulation of geometrically nonlinear Cosserat elasticity in the plane by following three different routes. First, we will follow the standard approach of formulating three dimensional Cosserat elasticity and restricting it to the plane, thereby recalling the first and second Cosserat problems, respectively. Next we will formulate an intrinsically two dimensional model. After stating the energy functional of our model, the equations of motion are rigorously derived using the calculus of variations. Our intrinsically two-dimensional formulation requires no reference to a theory in three dimensions. Within this setting we are able to introduce a new displacement vector which allows us to model planar chiral materials without the use of new constitutive relations. Our model, which is geometrically non-linear, yields equations very similar to those reported in~\cite{Liu20121907} when we assume small rotations and small displacements. However, due to our non-linear theory as a starting point, we find an additional contribution which naturally appears in the equations of motion.

A detailed list of symbols and our notation are given in the following.

\section*{Notation}

\begin{tabular}{ll}
  $\mathbbm{1}$ & identity matrix \\
  $\varphi$ & deformation vector in 3D and 2D\\
  $\phi$ & rotation angle in 3D \\
  $\mathbf{u}$ & displacement vector \\
  $\mathbf{a}$ & rotation vector \\
  $F=\nabla\varphi=I+\nabla\mathbf{u}$ & deformation gradient \\
  $F_{ij} = \delta_{ij} + \mathbf{u}_{i,j} = \delta_{ij} + \partial_j \mathbf{u}_{i}$ & deformation gradient in index notation\\
  $\overline{R} = \exp(\overline{A})$ & rotation matrix \\
  $\overline{A}$ & skew-symmetric matrix generating $\overline{R}$ \\
  $\epsilon$ & 2D Levi-Civita symbol, $\epsilon_{12} =1 = -\epsilon_{21}$, $\epsilon_{11} = \epsilon_{22} = 0$ \\
  $\overline{U} = \overline{R}^T F$ & non-symmetric stretch tensor, first Cosserat deformation tensor  \\
  $F = R\,U = \polar(F) U$ & classical polar decomposition \\
  $(\Curl M)_i = \varepsilon_{rs} \partial_r M_{is}$ & matrix Curl in 2D \\
  $(\Divv M)_i = \partial_s M_{is}$ & matrix Div in 2D \\
  $\sym M = (M+M^T)/2$ & symmetric part of matrix $M$\\
  $\skewp M  = (M-M^T)/2$ & skew-symmetric part of $M$\\
  $\dev M = M-\tr(M)\id/3$ & deviatoric or trace-free part of $M$ \\
  $\vartheta$ & rotation angle in 2D \\
  $\overline{R} = (\cos\vartheta) \id - (\sin\vartheta) \epsilon$ & rotation in 2D 
\end{tabular}
\\
Remark: we write the above tensors $\overline{R}$ and $\overline{U}$ with superposed bars in order to distinguish them from the factors $R$ and $U$ of the classical polar decomposition $F = R\,U$, in which $R = \polar(F)$ is orthogonal and $U$ is positive definite, symmetric and which is a standard notation in elasticity. We also note the standard relation $F^T F = (R\,U)^T (R\,U) = U^T R^T T U = U^T U = U^2$ so that $U=\sqrt{F^T F}$. \\
We will use the Frobenius scalar product defined as
\begin{align}
  \langle A,B \rangle = A:B = A_{ij}B_{ij}= \tr(AB^{T}) \,.
  \label{Frobenius}
\end{align}

\section{The Cosserat problem in the plane}

We are interested in studying the dynamical geometrically nonlinear Cosserat problem in the plane and this immediately poses the rather interesting question of how to formulate such a theory. On the one hand, one could simply start with three-dimensional Cosserat elasticity and consider an ansatz which reduces the equations to the planar case. This approach yields two different types of models which are often referred to as the first and second Cosserat planar model, see for instance~\cite{Joumaa20112896}. However, as Cosserat elasticity takes into account the possible microrotations of matter points, we must recall that rotations in the plane are very different to rotation in three dimensions, the former are an Abelian group while the latter are non-Abelian. Therefore, on the other hand, one could formulate an intrinsically two-dimensional model of Cosserat elasticity and it turns out that this model differs from the three-dimensional case restricted to the plane. 

\subsection{The first Cosserat planar problem}

The first Cosserat planar problem is defined by $\mathbf{u} = (u_1,u_2,0)$ and $\mathbf{a}=(0,0,a_3)$, where the vector $\mathbf{a}$ defines the axis of the Cosserat rotation. This means we can begin with a three-dimensional setup confined to the $x,y$-plane, with rotations about the $z$-axis only. In this case the deformation is given by $\varphi = (\varphi_1(x,y),\varphi_2(x,y),z)$ so that the deformation gradient $F$ reads
\begin{align}
  F = \nabla \varphi = 
  \begin{pmatrix}
    \varphi_{1,x} & \varphi_{1,y} & 0 \\
    \varphi_{2,x} & \varphi_{2,y} & 0 \\
    0 & 0 & 1
  \end{pmatrix} \,.
  \label{p1}
\end{align}
Next we consider the rotations in the $x,y$-plane. We call the angle of rotation $\phi=\phi(x,y)$ which yields
\begin{align}
  \overline{R} = 
  \begin{pmatrix}
    \cos(\phi) & -\sin(\phi) & 0 \\
    \sin(\phi) & \cos(\phi) & 0 \\
    0 & 0 & 1
  \end{pmatrix} \,.
  \label{p2}
\end{align}
Now we can compute $\overline{R}^T\! \Curl \overline{R}$, a useful curvature measure of the Cosserat theory~\cite{Neff_curl06}. Using the index notation, it can be written as
\begin{align}
  (\Curl \overline{R})_{ij} = \varepsilon_{jmn} \partial_m \overline{R}_{in} \,,
  \label{matrixcurl3d}
\end{align}
where $\varepsilon_{jmn}$ is the Levi-Civita symbol. For the given orthogonal matrix~(\ref{p2}) this is given by
\begin{align}
  \overline{R}^T\! \Curl \overline{R} = 
  \left(
  \begin{array}{cc|c}
    0 & 0 & -\phi_x \\
    0 & 0 & -\phi_y \\ \hline
    0 & 0 & 0
  \end{array}
  \right)\,.
  \label{p3}
\end{align}
At this point we already see the root cause of the problem when trying to formulate the desired theory. All the components of $\overline{R}^T\! \Curl \overline{R}$ restricted to the plane are zero. We also note that the components of this matrix are determined by the object $\grad \phi$, this means effectively a vector with two components which already points towards an intrinsically two-dimensional model.

A direct calculation establishes 
\begin{align}
  \dev (\overline{R}^T\! \Curl \overline{R}) &= 
  \frac{1}{2}\begin{pmatrix}
    0 & 0 & -\phi_x \\
    0 & 0 & -\phi_y \\
    -\phi_x & -\phi_y & 0
  \end{pmatrix} \,, 
  \nonumber \\
  \skewp (\overline{R}^T\! \Curl \overline{R}) &= 
  \frac{1}{2}\begin{pmatrix}
    0 & 0 & -\phi_x \\
    0 & 0 & -\phi_y \\
    \phi_x & \phi_y & 0
  \end{pmatrix} \,, 
  \nonumber \\
  \tr (\overline{R}^T\! \Curl \overline{R}) &= 0 \,.
  \label{p4}
\end{align}
The important issue at this point is that there are no planar contributions in any of the irreducible parts of $\overline{R}^T\! \Curl \overline{R}$ in (\ref{p4}).

Also, we can compute the non-symmetric stretch tensor which gives
\begin{align}
  \overline{U} = \overline{R}^T F = 
  \begin{pmatrix}
    \varphi_{1,x}\cos(\phi) + \varphi_{2,x}\sin(\phi) & \varphi_{1,y}\cos(\phi) + \varphi_{2,y}\sin(\phi) & 0 \\
    \varphi_{2,x}\cos(\phi) - \varphi_{1,x}\sin(\phi) & \varphi_{2,y}\cos(\phi) - \varphi_{1,y}\sin(\phi) & 0 \\
    0 & 0 & 1
  \end{pmatrix} \,.
  \label{p5}
\end{align}
This implies that the irreducible parts of $\overline{U}=\overline{R}^T F$ and $\overline{R}^T\! \Curl \overline{R}$ are orthogonal in the sense that
\begin{align}
  \tr (\overline{R}^T\! \Curl \overline{R}) \tr(\overline{R}^T F) &= 0 \,,
  \nonumber \\
  \langle \dev (\overline{R}^T\! \Curl \overline{R}), \dev(\overline{R}^T F) \rangle &= 0 \,,
  \nonumber \\
  \langle \skewp (\overline{R}^T\! \Curl \overline{R}), \skewp(\overline{R}^T F) \rangle &= 0 \,.
  \label{p6}
\end{align}
This implies that we cannot construct interaction terms between the displacements and the microrotations. Interaction terms of the above form were considered in~\cite{boehmer2013rota,Boehmer2016158}. These terms allow for a natural coupling between elastic deformations and the microrotations which, for instance, gives rise to soliton solutions, see~\cite{Boehmer2016158}. 

Let us also remark that in case of the finite theory one may as well consider the irreducible components of $F^T F$ rather and $\overline{R}^T F$. The form of $F$ given in (\ref{p1}) implies that $F^T F$ will be the same form as $\overline{U}$ given by (\ref{p5}). Hence, all inner product considered in (\ref{p6}) would also vanish which implies that it is also not possible to construct interaction terms in the finite theory. 

However, one can still construct interaction term, albeit less natural ones. Consider the Frobenius norm of the dislocation density tensor
\begin{align}
  \|(\overline{R}^T\! \Curl \overline{R})\|^2 = \phi_x^2 + \phi_y^2 \,,
  \label{p7}
\end{align}
which agrees with the vector norm of $\grad \phi$. Hence, the only option to construct an interaction term would be to consider an expression of the form 
\begin{align}
  V_{\rm interaction} \propto \|(\overline{R}^T\! \Curl \overline{R})\| \, \tr(\overline{R}^T F) \,.
  \label{p8}
\end{align}
An interaction of this type is quite unnatural since we are not considering the inner products of objects in the same irreducible spaces which is what led to (\ref{p6}). From a more mathematical point of view, the presence of the square root in (\ref{p8}) might cause differentiability issues when the orthogonal matrix $\overline{R}$ approaches a constant rotation. 

It is also clear from (\ref{p5}) that the polar part of any deformation tensor restricted to the plane will be of the same form as the rotation matrix (\ref{p2}) which means we could, in principle, construct the Cosserat couple term which contains the term $\overline{R}^T\negmedspace\polar(F)$.

\subsection{The second Cosserat planar problem}

The second Cosserat planar problem is defined by $\mathbf{u} = (0,0,u_3)$ and $\mathbf{a}=(a_1,a_2,0)$. This means $\varphi = (x,y,\varphi_3(t,x,y))$, so that elastic displacements are only allowed along the direction perpendicular to the plane. The rotations can be about the $x$-axis and the $y$-axis in this case. While this is mathematically well-defined, this is less well motivated than the first Cosserat problem, from a practical point of view.

The deformation gradient of the second Cosserat planar model is given by
\begin{align}
  F = \nabla \varphi = 
  \begin{pmatrix}
    1 & 0 & 0 \\
    0 & 1 & 0 \\
    \varphi_{3,x} & \varphi_{3,y} & 1
  \end{pmatrix} \,.
  \label{sec_p1}
\end{align}
Next we consider the rotations about the $x$ and about the $y$ axes. We call the respective angles of rotation $\alpha=\alpha(t,x,y)$ and $\beta=\beta(t,x,y)$ so that
\begin{align}
  \overline{R} = 
  \begin{pmatrix}
    \alpha^2/\ell^2 + \cos(\ell)\beta^2/\ell^2 & 
    (1-\cos(\ell))\alpha\beta/\ell^2 & 
    \sin(\ell) \beta/\ell \\
    (1-\cos(\ell))\alpha\beta/\ell^2 & 
    \cos(\ell)\alpha^2/\ell^2 + \beta^2/\ell^2 & 
    -\sin(\ell) \alpha/\ell\\
    -\sin(\ell) \beta/\ell & 
    \sin(\ell) \alpha/\ell & 
    \cos(\ell)
  \end{pmatrix}
  \,,
  \label{sec_p2}
\end{align}
where $\ell = \sqrt{\alpha^2+\beta^2}$. Due to the more complicated structure of the rotation matrix, the second Cosserat planar problem differs from the first problem substantially. This is due to our geometrically non-linear setup which allows for large rotations. The explicit form of $\overline{R}^T\! \Curl \overline{R}$ is rather involved and therefore we will note state it explicitly. It suffices to mention that the only vanishing components are $(\overline{R}^T\! \Curl \overline{R})_{31} = (\overline{R}^T\! \Curl \overline{R})_{32} = 0$. It is instructive, however, to consider this quantity assuming small rotations $\alpha,\beta \ll 1$ in which case one finds
\begin{align}
  \overline{R}^T\! \Curl \overline{R} \simeq 
  \begin{pmatrix}
    \beta_y & -\beta_x & 0 \\
    -\alpha_y & \alpha_x & 0 \\
    0 & 0 & \alpha_x + \beta_y
  \end{pmatrix} \,.
  \label{sec_p3}
\end{align}
One interesting aspect of this equation is the presence of the $(zz)$ components. When linear Cosserat elasticity is considered, as in~\cite{Joumaa20112896}, then the curvature tensor does not have this component and is indeed restricted to the plane. In the non-linear setting this is no longer the case which also motivates a different approach to the planar case. 

It is clear that the stretch tensor $\overline{U} = \overline{R}^T F$ is not orthogonal to any of the components of $\overline{R}^T\! \Curl \overline{R}$. Therefore the aforementioned coupling terms could in principle be constructed. However, this construction appears to be quite unnatural when compared to the first Cosserat problem.

The conceptual problem of using either the first or the second Cosserat planar model is simply that both theories differ considerably, and also differ from the non-linear theory. This motivates us to formulate an intrinsically two-dimensional model which does not refer to the three-dimensional setting altogether. 

\section{Intrinsic planar model}

\subsection{Basic quantities}

In order to formulate an intrinsically two-dimensional model we simply begin with the two-dimensional deformation vector $\varphi = (\varphi_1,\varphi_2)$ so that the deformation gradient reads
\begin{align}
  F = \nabla \varphi = 
  \begin{pmatrix}
    \varphi_{1,x} & \varphi_{1,y} \\
    \varphi_{2,x} & \varphi_{2,y}
  \end{pmatrix} \,.
\end{align}
Similar to rotations about the $z$-axis in the first Cosserat planar model, we consider a two-dimensional rotation matrix where, as before, we call the angle $\vartheta=\vartheta(x,y)$ so that
\begin{align}
  \overline{R} = 
  \begin{pmatrix}
    \cos(\vartheta) & -\sin(\vartheta) \\
    \sin(\vartheta) & \cos(\vartheta)
  \end{pmatrix} \,.
  \label{p9}
\end{align}
These definitions are purely two-dimensional and do not require the higher dimensional setting. However, as somewhat expected, the components of the intrinsic deformation gradient are identical to the planar components of the corresponding three-dimensional one, see eq.~(\ref{p1}). Likewise, the intrinsic rotation matrix has the planar components of the three-dimensional rotation matrix (\ref{p2}) of the first Cosserat planar problem.

The first object which requires a more careful approach is the matrix Curl of the curvature measure. In three dimensions the matrix Curl requires the object $\varepsilon_{jmn}$, see eq.~(\ref{matrixcurl3d}). However, the Levi-Civita tensor $\varepsilon_{jmn}$ is a three-dimensional object which has no geometrical meaning in any dimension other than three. Therefore, one has to change this definition and adopt it to the planar case.

Let us begin recalling the lesser-known fact that the vector curl is sometimes introduced in two dimensions to give a scalar quantity. Consequently, one would expect the two-dimensional matrix Curl to give a vector. It turns out that this follows quite naturally when the Levi-Civita symbol in two dimensions is considered. Namely, it has two indices $\epsilon_{ij}$, instead of $\varepsilon_{ijk}$ in three dimensions. Hence, the contraction of the partial derivative of a matrix with the Levi-Civita gives a vector. This leads us to define the two-dimensional matrix curl $\overline{R}^T\negmedspace \Curl \overline{R}$ in the following way
\begin{align}
  (\Curl \overline{R})_i := \epsilon_{rs} \partial_r \overline{R}_{is} \,.
  \label{p10}
\end{align}
This matrix Curl maps matrices to vectors as expected. 

This definition is quite natural in the present context as can be seen by computing the quantity $\overline{R}^T\! \Curl \overline{R}$ which then becomes
\begin{align}
  \overline{R}^T\! \Curl \overline{R} =
  \begin{pmatrix}
    \vartheta_x \\
    \vartheta_y
  \end{pmatrix} = \grad \vartheta \,.
  \label{p11}
\end{align}
This result contains the two non-vanishing components of the three-dimensional object $\overline{R}^T\! \Curl \overline{R}$, compare with eq.~(\ref{p3}).  

Now, we are faced with a similar problem as before, we cannot simply couple the two-dimensional vector (the two-dimensional Cosserat curvature measure) to the deformation gradient which is, of course, a $2 \times 2$ matrix. However, we can proceed as follows. Let us write any vector $\mathbf{v}$ as $\mathbf{v} = |\mathbf{v}|\, \hat{\mathbf{v}}$, which we can view as the irreducible decomposition of the vector into a scalar $|\mathbf{v}|$ and a direction $\hat{\mathbf{v}}$ with $|\hat{\mathbf{v}}|=1$. Applied to the vector $\overline{R}^T\! \Curl \overline{R}$, we can consider the scalar quantity $\|\overline{R}^T\! \Curl \overline{R}\|$ which can be coupled to the trace part of the deformation gradient. Hence, we can consider the term 
\begin{align}
  V_{\rm interaction} \propto \tr(\overline{R}^T F) \, \|\overline{R}^T\negmedspace \Curl \overline{R}\|\,,
  \label{p12}
\end{align}
which in the two-dimensional setting is a natural choice. We recall that in the three-dimensional case we could write a very similar term, namely (\ref{p8}), however, it was not well justified in that setting. 

The main point to note is that modelling planar material requires, independently of the approach, two geometrically different quantities. Matrix valued objects which describe the elastic deformations and vectors which describe the rotations about the axis perpendicular to the material. Any model, which attempts at introducing interactions between these two quantities, needs to address the principal issue of how such couplings can be achieved.

Since $\overline{R}$ is still an orthogonal matrix, it appears to be best motivated to consider the coupling based on $\overline{R}^T\negmedspace \polar(F)$ which does not involve conceptual problems in either dimensions two or three. 

The Cosserat model is cast in a variational framework on the reference configuration. The dynamical equations follow from a generalized Hamiltonian principle. A replacement of the first Piola-Kirchhoff tensor of classical nonlinear elasticity is easily seen to be $\partial V/\partial F_{ij}$ with $V$ being the energy function of the model which is considered.

\subsection{Elastic energy and curvature energy}

We write the energy functional for the elastic deformation as
\begin{align}
  V_{\rm elastic}(F,\overline{R}) = 
  \mu\, \| \sym \overline{R}^T F - \id\|^2 + 
  \frac{\lambda}{2}\left(\tr(\sym(\overline{R}^T F) - \id)\right)^2 \,,
  \label{Velas}
\end{align}
where $\lambda$ and $\mu$ are the standard elastic Lam\'{e} parameters in two dimensions. For the dynamical treatment, we will need to subtract kinetic energy which we assume to be of the form $\rho/2 \, \|\dot{\varphi}\|^2$.

The three dimensional energy functional of the micro-rotations $V_{\rm rotational}$ is based on the energy functional containing $\|\overline{R}^T \Curl \overline{R}\|^2_{\mathbb{R}^{3 \times 3}}$. It is natural to consider the same functional form in two dimensions using the matrix Curl defined in $\mathbb{R}^2$, as in eq.~(\ref{p10}).

Using the previously stated equation (\ref{p11}), we write the two-dimensional curvature energy as
\begin{align}
  V_{\rm curvature}(\nabla\vartheta) = 
  \mu L_c^2\, \|\overline{R}^T\negmedspace \Curl \overline{R}\|_{\mathbb{R}^2}^2 = 
  \mu L_c^2\, \|\grad \vartheta\|_{\mathbb{R}^2}^2 \,,
  \label{Vcurv}
\end{align}
where $L_c$ is the characteristic length and $\mu$ is the shear modulus from above. We emphasise that this (vector) norm is computed in two dimensions by using the subscript $\mathbb{R}^2$, this is to avoid confusion with the 3D (matrix) norm in $\mathbb{R}^{3 \times 3}$. The simple form of the energy when expressed in vector form is in fact expected as rotations in the plane are characterised by only one angle $\vartheta$.

In order to study the dynamical problem, we will need to subtract kinetic energy which we assume to be of the form $\rho_{\rm rot} \|\dot{\vartheta}\|^2$ where $\rho_{\rm rot}$ is the scalar rotational density. One could also choose the kinetic energy as $\tr(\dot{\overline{R}}^T \dot{\overline{R}})$ as this is somewhat more natural, however, a direct calculation gives that $\tr(\dot{\overline{R}}^T \dot{\overline{R}}) = 2 \|\dot{\vartheta}\|^2$ which means that they only differ by a factor of two.  

\subsection{Interaction and coupling terms}

Next, we wish to introduce a coupling between the elastic displacements and the micro-rotations. In order to do so, we will `couple' the irreducible parts of the elastic deformation with the micro-rotations. This gives 
\begin{align}
  V_{\rm interaction}(F,\overline{R}) =
  \mu L_c \chi\, \|\overline{R}^T \Curl \overline{R}\| \, \tr(\overline{R}^T F) \,,
  \label{Vint}
\end{align}
where $\chi$ is the coupling constant which was first introduced in~\cite{boehmer2013rota,Boehmer2016158}.

Finally, we will consider the Cosserat couple term we assume to be given by
\begin{align}
  V_{\rm coupling}(F,\overline{R}) = 
  \mu_c \| \overline{R}^T \polar(F) - \id \|^2 \,,
  \label{Vcoupagain}
\end{align}
where $\mu_c \geq 0$ is the Cosserat couple modulus. Alternatively, one can consider the coupling
\begin{align}
  V_{\rm coupling(2)}(F,\overline{R}) = 
  \mu_c \| \skewp (\overline{R}^T F - \id)\|^2 \,,
  \label{Vcoupagain2}
\end{align}
which induces the same linear response as the coupling containing the polar part.

The complete static model is therefore given by
\begin{align}
  V = 
  {}&{} 
  V_{\rm elastic} + V_{\rm curvature} + 
  V_{\rm interaction} + V_{\rm coupling} 
  \nonumber \\
  = {}&{} 
  \mu\, \| \sym (\overline{R}^T F - \id)\|_{\mathbb{R}^{2 \times 2}}^2 + 
  \frac{\lambda}{2}\left(\tr(\sym(\overline{R}^T F - \id))\right)^2 +
  \mu L_c^2\, \|\overline{R}^T \Curl \overline{R}\|_{\mathbb{R}^2}^2 
  \nonumber \\ 
  {}&{} +
  \mu L_c \chi\, \|\overline{R}^T \Curl \overline{R}\|_{\mathbb{R}^2} \, \tr(\overline{R}^T F) +
  \mu_c \| \overline{R}^T\! \polar(F) - \id \|_{\mathbb{R}^{2 \times 2}}^2 \,.
  \label{fullenergy}
\end{align}
Where necessary we indicated the space over which the norm has to be computed. This is only to clarify this expression as it should be clear from the context which is the appropriate space. 

\subsection{Modelling to chiral lattices}

The proposed intrinsic model can be applied to chiral lattices. It turns out that considering small elastic displacements and small microrotations leads to a model very similar to the one studied in \cite{Liu20121907} with an additional term contributing to the dynamics of the microrotations. In order to allow for chirality to be incorporated into our approach we define the quantity
\begin{align}
  u^{\ast}_{i} = \varepsilon_{ij} u_j\,.
\end{align}
In the three dimensional setting one could not construct a term like this as the object $\varepsilon_{ij}$ is intrinsically two dimensional. The corresponding term in three dimensions is $\varepsilon_{ijk}$ which would map the vector $u_k$ to a matrix. The two-dimensional Levi-Civita symbol $\varepsilon_{ij}$ is identical to our rotation matrix $\overline{R}$ for $\vartheta=-\pi/2$. This means we can write
\begin{align}
  \varepsilon =
  \begin{pmatrix} 0 & 1 \\ -1 & 0 \end{pmatrix}\,.
\end{align}
Thus, the vector $\mathbf{u}^\ast$ is the vector $\mathbf{u}$ rotated by $90^\circ$ in the counter-clockwise direction and is given by
\begin{align}
  \mathbf{u}^{\ast} = \varepsilon \mathbf{u} =
  \begin{pmatrix} 0 & 1 \\ -1 & 0 \end{pmatrix}
  \begin{pmatrix} u_1 \\ u_2 \end{pmatrix} =
  \begin{pmatrix} u_2 \\ -u_1 \end{pmatrix} \,.
\end{align}

Now we define the corresponding deformation gradient of the rotated vector
\begin{align}
  F^{\ast} = \id + \nabla\mathbf{u}^\ast\,,
  \label{fstar}
\end{align}
so that we can define a corresponding elastic energy
\begin{align}
  V^\ast_{\rm elastic}(F^\ast,\overline{R}) = 
  \mu^\ast\, \| \sym \overline{R}^T F^\ast - \id\|^2 + 
  \frac{\lambda^\ast}{2}\left(\tr(\sym(\overline{R}^T F^\ast) - \id)\right)^2 +
  \mu_c^\ast \| \skewp (\overline{R}^T F^\ast - \id)\|^2\,.
  \label{Velasstar}
\end{align}
The use of $F^\ast$ in this new elastic energy is the only difference to~(\ref{Velas}). This means we are introducing a new elastic energy which depends on the rotated vector $\mathbf{u}^\ast$. Let us emphasise again, that an analogous constructing in three dimensions cannot be achieved. 

Due to these new terms based on $u^\ast$, we can also define the following mixing terms
\begin{align}
  V_{\rm mixing} = 
  m_1 \tr \Bigl[(\sym(\overline{R}^T F^\ast) - \id)^T(\sym(\overline{R}^T F) - \id)\Bigr] +
  m_2 \tr(\overline{R}^T F^\ast - \id) \tr(\overline{R}^T F - \id)\,,
  \label{Vmix}
\end{align}
where we note that we could also introduce a third mixing term of the form 
\begin{align*}
  m_3 \tr \Bigl[(\skewp(\overline{R}^T F^\ast - \id))^T(\skewp(\overline{R}^T F - \id))\Bigr] \,.
\end{align*} 
As this term will not be required in what follows, we will neglect this contribution henceforth.

Putting together everything that is needed to model chiral effects leads to the energy functional
\begin{align}
  V_{\rm chiral}(F,F^\ast,\overline{R}) = V_{\rm curvature} +
  V_{\rm elastic} + V^\ast_{\rm elastic} + V_{\rm mixing} + V_{\rm coupling(2)} \,.
  \label{chiral}
\end{align}

\section{Equations of motion}

In the following subsections we will state the equations of motion of our model which are derived using the calculus of variations. While most of this is fairly standard from a mathematical point of view, it is instructive to provide enough detail of this derivation, most of which can be found in Appendix~\ref{app:vari}. We work predominantly in the matrix notation and we also need to consider variations with respect to the polar part of the deformation gradient, which is a non-standard result. Also our use of matrix curls and matrix divergences requires a careful treatment which is shown. 

\subsection{Field equations without chiral terms}

The complete variational energy functional of the nonlinear Cosserat micropolar theory in 2D will be the sum of each variational term, so we have
\begin{align}
  \delta V =\delta V_{\rm curvature} + \delta V_{\rm elastic} + \delta V_{\rm interaction} + \delta V_{\rm coupling} \,. 
  \label{Vcomplete}
\end{align}
Rotations in two dimensions depend only on one angle and one can verify that 
\begin{align}
  \delta \overline{R} = -(\cos(\vartheta)\epsilon+\sin(\vartheta)I) \,\delta\vartheta =
  -\epsilon\overline{R}\, \delta \vartheta \,,
  \label{deltaRvartheta}
\end{align}
where $\epsilon$ is the 2D Levi-Civita matrix. While we computed the variations with respect to $\delta F$, we are in fact interested in the variations with respect to the displacements $\delta u$. So, let $M$ be an arbitrary matrix then we can easily find the relation between $\delta F$ and $\delta u$ which is given by
\begin{align}
  \langle A,\delta F \rangle = A:\delta F & = -(\Divv M)\, \delta u \,,
  \label{deltaFu}
\end{align}
or in other words, we need to integrate by parts once more to arrive at the equations of motions in the displacements.

We can now use (\ref{deltaRvartheta}) and (\ref{deltaFu}) to rewrite the variations of Appendix~\ref{app:vari} as follows
\begin{align}
  \delta V_{\rm elastic} = {}&-
  \Big(\mu F\overline{R}^{T}F - 2(\mu+\lambda)F + \lambda \tr(\overline{R}^{T}F)F\Big):
  (\epsilon \overline{R})\,\delta\vartheta
  \nonumber \\ {}&-
  \Divv\Big[
    \mu(\overline{R}F^{T}\overline{R}+F) - 2(\mu+\lambda)\overline{R} + 
    \lambda\tr(\overline{R}^{T}F)\overline{R}
  \Big] \, \delta u \,,
  \label{var1} \\
  \delta V_{\rm curvature} = {}&-2 \mu L_c^2 \divv(\grad\vartheta) \, \delta\vartheta \,,
  \label{var2} \\
  \delta V_{\rm interaction} = {}&-
  \mu L_c \chi \Big(\divv\Big[\tr(\overline{R}^{T}F)\frac{\grad\vartheta}{\|\grad\vartheta\|}\Big] +
  \|(\grad\vartheta\|F\Big): 
  (\epsilon \overline{R}) \,\delta\vartheta 
  \nonumber \\ {}&-
  \mu L_c \chi \Divv\Big[\|(\grad\vartheta)\|\overline{R}\Big]\,\delta u \,,
  \label{va3} \\
  \delta V_{\rm coupling} = {}&
  2\mu_{c} \polar(F):(\epsilon \overline{R}) \,\delta\vartheta +
  2\mu_{c}\Divv\Big[\frac{1}{\tr(U)}(\overline{R}-\polar(F)\overline{R}^{T}\polar(F))\Big]\, \delta u \,.
\end{align}

\subsection{Fully nonlinear equations without chiral terms}

The variations with respect to our two dynamical variables, the vector $u$ and the scalar $\vartheta$ give two Euler-Lagrange equations. The equation of motion including kinetic energy for the displacement vector $\mathbf{u}$ are now written in their final form which is given by
\begin{multline}
  \rho\, u_{tt} = \Divv
  \Big[ 
    2 \mu \overline{R}\, \sym(\overline{R}^T F) +
    \lambda\, \tr(\overline{R}^{T}F)\overline{R} -
    2(\mu+\lambda)\overline{R} \\ + 
    \frac{4\mu_{c}}{\tr(\sqrt{F^T F})} \polar(F) \skewp(\overline{R}^{T}\!\polar(F)) + 
    \mu L_c \chi\, \|(\overline{R}^T\! \Curl \overline{R})\| \, \overline{R} 
  \Big] \,.
  \label{fieldeq2a}
\end{multline}
Here we used the useful identities $2\polar(F) \skewp(\overline{R}^{T}\!\polar(F)) = -\overline{R}+\polar(F)\overline{R}^{T}\!\polar(F)$ and moreover $2\overline{R}\,\sym(\overline{R}^{T}F)=\overline{R}F^{T}\overline{R}+F$.

Before stating the equation of motion for the rotation, let us have a closer look at the various terms. The first line corresponds to the equations of nonlinear elasticity and the second line contains the interaction term and the Cosserat couple term. 

Variations with respect to the rotation yield the following equation of motion
\begin{multline}
  \rho_{\rm rot}\, \vartheta_{tt} = \mu L_c^2 \divv(\overline{R}^T\! \Curl \overline{R}) -
  (\mu+\lambda)\tr(\epsilon\, \overline{R}^T F) +
  \frac{\mu}{2} \tr(\epsilon\, (\overline{R}^{T}F)^2) +
  \frac{\lambda}{2} \tr(\overline{R}^{T}F) \tr(\epsilon\, \overline{R}^{T}F) \\ +
  \frac{1}{2} \mu L_c \chi
  \bigg\{
  \divv \Big[\frac{(\overline{R}^T\! \Curl \overline{R})}{\|(\overline{R}^T\! \Curl \overline{R})\|}\tr(\overline{R}^{T}F)\Big] +
  \|(\overline{R}^T\! \Curl \overline{R})\|\tr(\epsilon\, \overline{R}^{T} F) 
  \bigg\} -
  \mu_{c} \tr(\epsilon\, \overline{R}^T\! \polar(F)) \,.
  \label{fieldeq2b}
\end{multline}
The most complicated term comes from the interaction term and in particular the presence of the square-root when we work with the norm of $\overline{R}^T\negmedspace \Curl \overline{R}$. One has to be careful with this term as the square root is not differentiable at the origin. Due to the nonlinear nature of these equations it is very difficult finding explicit solutions, or making generic statements about such solutions.

\subsection{Fully nonlinear equations of the chiral model}

The complete variational energy functional of the nonlinear Cosserat micropolar theory with chiral terms is given by
\begin{align}
  \delta V = \delta V_{\rm curvature} + \delta V_{\rm elastic} + \delta V^\ast_{\rm elastic} + \delta V_{\rm mixing} + \delta V_{\rm coupling(2)} \,.
  \label{Vchiral2}
\end{align}
The equations of motion of the chiral model for the displacement $u$ are given by
\begin{align}
  \rho\, u_{tt}  = {}&{} \Divv
  \Big[ 
    2 \mu \overline{R}\, \sym(\overline{R}^T F) +
    \lambda\, \tr(\overline{R}^{T}F)\overline{R} -
    2(\mu+\lambda)\overline{R} + 
    \mu_{c}(F-\overline{R}F^{T}\overline{R}) 
    \Big]
  \nonumber \\ {}&{} -\Divv
  \Big[
  \mu^{\ast}(\overline{R}(F^{\ast})^{T}\overline{R}+F^{\ast}) -
  2(\mu^{\ast}+\lambda^{\ast})\overline{R} + 
  \lambda^{\ast} \,\tr(\overline{R}^{T}F^{\ast})\overline{R}
  + \mu_c^{\ast} (\overline{R}(F^{\ast})^{T}\overline{R}-F^{\ast})
  \Big]:\epsilon
  \nonumber\\ {}&{} -\Divv
  \Bigl[
    \frac{1}{2}m_{1}
    \Big(\epsilon^{T}\overline{R}F^{T}\overline{R}+\overline{R}(F^{\ast})^{T}\overline{R}+F:\epsilon+F^{\ast}-2(\overline{R}:\epsilon+\overline{R})
    \Big)
    \nonumber \\ {}&{} +
    m_{2} \Big(\tr(\overline{R}^{T}F^{\ast})R+\tr(\overline{R}^{T}F)(R:\epsilon)-(\overline{R}:\epsilon+\overline{R})\Big)
    \nonumber\\ {}&{} +
    \frac{1}{2}m_{3}\Big(F:\epsilon+F^{\ast} -(F\overline{R}^{T}\epsilon^{T}F+\epsilon^{T}F\overline{R}F)\Big)
    \Big]\,.
\end{align}
Additionally, the equation of motion for $\vartheta_{tt} $ is given by
\begin{align}
  \rho_{\rm rot}\, \vartheta_{tt} = {}& {}
  \mu L_c^2 \divv(\overline{R}^T\! \Curl \overline{R}) -
  (\mu+\lambda)\tr(\epsilon\, \overline{R}^T F) +
  \frac{\mu}{2} \tr(\epsilon\, (\overline{R}^{T}F)^2)
  \nonumber \\ {}&{} +
  \frac{\lambda}{2} \tr(\overline{R}^{T}F) \tr(\epsilon\, \overline{R}^{T}F) +
  \mu_{c} \tr(\epsilon\, \overline{R}^T F \overline{R}^T)
  \nonumber \\ {}&{}+
  \Big[
    \mu^{\ast} F^{\ast}\overline{R}^{T}F^{\ast} -
    2(\mu^{\ast}+\lambda^{\ast})F^{\ast} + 
    \lambda^{\ast} \, \tr(\overline{R}^{T}F^{\ast})F^{\ast} +
    \mu_c^{\ast} F^{\ast}\overline{R}^{T}F^{\ast}
    \nonumber\\ {}&{} +
    \frac{1}{2}m_{1}
    \Big(F\overline{R}^{T}\epsilon^{T}F+\epsilon^{T}F\overline{R}F-2(F+F^{\ast})\Big) +
    m_{2}\Big(\tr(\overline{R}^{T}F^{\ast})F+\tr(\overline{R}^{T}F)F^{\ast}-(F+F^{\ast})\Big)
    \nonumber\\ {}&{} -
    \frac{1}{2}m_{3}\Big(F\overline{R}^{T}\epsilon^{T}F+\epsilon^{T}F\overline{R}F\Big)\Big]:(-\epsilon \overline{R})\,.
\end{align}
Before having a closer look at these equations and discussing applications to chiral lattices, we will briefly state two special solutions to the equations without chiral terms.

\subsection{Special solution -- no displacements, static and homogeneous microrotations}

One of the simplest possible solutions can be sought in the form $u \equiv 0$ and $\vartheta = \vartheta_0$ where $\vartheta_0$ is a constant angle. Using this ansatz, the equations for the displacements (\ref{fieldeq2a}) are identically satisfied. Equation (\ref{fieldeq2b}) for the microrotations reduces to the simple equation
\begin{align}
  \left(\lambda+\mu+\mu_c-(\lambda+\mu)\cos(\vartheta_0)\right)\sin(\vartheta_0) = 0 \,.
  \label{hom1}
\end{align}
This equation has two somewhat trivial solutions when $\vartheta_0=0$ or $\vartheta=\pi$. These correspond to all the oriented material points to be aligned horizontally. Interestingly, there are more solutions and their properties depend on the value of the Cosserat couple modulus $\mu_c$. If $\mu_c=0$, then (\ref{hom1}) is also satisfied by $\vartheta_0=\pi/2$ or $\vartheta_0=3\pi/2$ which corresponds to all the oriented material points to be aligned vertically.

On the other hand, if $\mu_c > 0$, then we find
\begin{align}
  \cos(\vartheta_0) = 1 + \frac{\mu_c}{\lambda+\mu} \,,
  \label{hom2}
\end{align}
for which a solution exits provided that $\lambda + \mu \leq 0$. This, however, contradicts the standard assumptions of linear elasticity. These results appear to be consisted with our expectations of the theory.

\subsection{Special solution -- no displacements, static and homogeneous microrotations with chiral terms}

Next, let us consider $u \equiv 0$ and $\vartheta = \vartheta_0$ in the equations with chiral terms. The equations for the displacements (\ref{fieldeq2a}) are again identically satisfied. Equation (\ref{fieldeq2b}) for the microrotations reduces to
\begin{multline}
  \Bigl[-m_1 - 2 m_2 - \lambda + \lambda^\ast - \mu - \mu_{c1} + \mu^\ast \\+ 
  (m_1 + 2 m_2 - m_3 - \mu_c - \mu_c^\ast + \lambda + \lambda^\ast + \mu + \mu^\ast) 
  \cos(\phi_0)\Bigr] \sin(\phi_0) = 0 \,,
  \label{hom21}
\end{multline}
where we also include the chiral and the mixing terms. As before, this equation has two trivial solutions when $\vartheta_0=0$ or $\vartheta=\pi$. Lastly,
\begin{align}
  \cos(\vartheta_0) = 1 + 
  \frac{\mu_{c1} + \mu_c + \mu_c^\ast - 2 \lambda^\ast - 2\mu^\ast + m_3}
       {\lambda + \lambda^\ast + \mu + \mu^\ast - \mu_c - \mu_c^\ast + m_1 + 2 m_2 - m_3} \,,
  \label{hom22}
\end{align}
for which a solution exits provided that the fraction is between $0$ and $-2$. Due to the extra parameters present in this theory there is no \textit{a priori} reason for this equation to have no solutions.

\section{Applications to chiral lattices}

Finally, we are able to demonstrate that our intrinsic two-dimensional model is applicable to chiral lattices. In doing so it is shown that it may not be necessary to begin with a three-dimensional theory and reduce it to the plane.

\subsection{Equations of motion}

By assuming small microrotations and small elastic displacements, we arrive at the following set of equations where we re-scaled $\rho_{\rm rot} = \varrho_{\rm rot}/4$ to match the factors used in~\cite{Liu20121907} and also introduce $\vartheta = -\phi$. This gives
\begin{align}
  \rho \frac{\partial^2 u_1}{\partial t^2} = 
  {}& (\lambda + 2\mu + \mu^\ast + \mu_c^\ast)\, u_{1,xx} +
  (\mu + \mu_c + \lambda^\ast + 2\mu^\ast)\, u_{1,yy} +
  (\lambda + \mu - \mu_c - \lambda^\ast - \mu^\ast + \mu_c^\ast)\, u_{2,xy} 
  \nonumber \\ &+
  (m_1/2+m_2)(-2u_{1,xy} + u_{2,xx} - u_{2,yy}) -
  2(2\lambda^\ast + 2\mu^\ast - \mu_c^\ast) \phi_x +2\mu_c\, \phi_y \,,
  \label{eom1}
\end{align}
and the second equation
\begin{align}
  \rho \frac{\partial^2 u_2}{\partial t^2} = 
  {}& (\mu + \mu_c + \lambda^\ast + 2\mu^\ast)\, u_{2,xx} +
  (\lambda + 2\mu + \mu^\ast + \mu_c^\ast)\, u_{2,yy} +
  (\lambda + \mu - \mu_c - \lambda^\ast - \mu^\ast + \mu_c^\ast)\, u_{1,xy} 
  \nonumber \\ &+
  (m_1/2+m_2)(u_{1,xx} - u_{1,yy} + 2u_{2,xy}) -
  2\mu_c\, \phi_x - 2(2\lambda^\ast + 2\mu^\ast - \mu_c^\ast) \phi_y \,.
  \label{eom2}
\end{align}
Lastly, the equation for the microrotations is given by
\begin{align}
  \varrho_{\rm rot} \frac{\partial^2 \phi}{\partial t^2} = 
  {}& 2 d_1 (\phi_{xx} + \phi_{yy}) + 
  4(2\lambda^\ast+2\mu^\ast-\mu_c^\ast-\mu_c) \phi +
  2\mu_c(u_{2,x}-u_{1,y})
  \nonumber \\ &-
  2(\mu_c^\ast-2\lambda^\ast-2\mu^\ast)(u_{1,x} + u_{2,y}) \,.
  \label{eom3}
\end{align}
Setting $2d_1=\gamma$, $\mu^\ast=-\mu_c^\ast=A$, $\lambda^\ast=-2A$, and $m_1/2+m_2=-A$, we recover the equations reported in \cite[see eqs. (17)]{Liu20121907}. The only difference is an additional term in the rotational equation, namely this becomes
\begin{align}
  \varrho_{\rm rot} \frac{\partial^2 \phi}{\partial t^2} = 
  {}& \gamma (\phi_{xx} + \phi_{yy}) - 
  4(\mu_c+A) \phi +
  2\mu_c(u_{2,x}-u_{1,y}) - 2A(u_{1,x} + u_{2,y}) \,,
  \label{eom3b}
\end{align}
where the term $-4A\, \phi$ is the additional contribution. We believe that this term is present due to our starting point being the fully non-linear model. Consequently, when studying plane wave solutions of this modified model, we expect some changes with respect to the wave speeds and the ratio of the amplitudes of the elastic waves as will be discussed henceforth. It appears that our model can also be used in the context of tetrachiral lattices, see the equations of motion reported in \cite{Chen20130734}.

\subsection{Plane wave solutions for the chiral model}

\begin{figure}[htb!]
  \centering
  \includegraphics[width=0.7\textwidth]{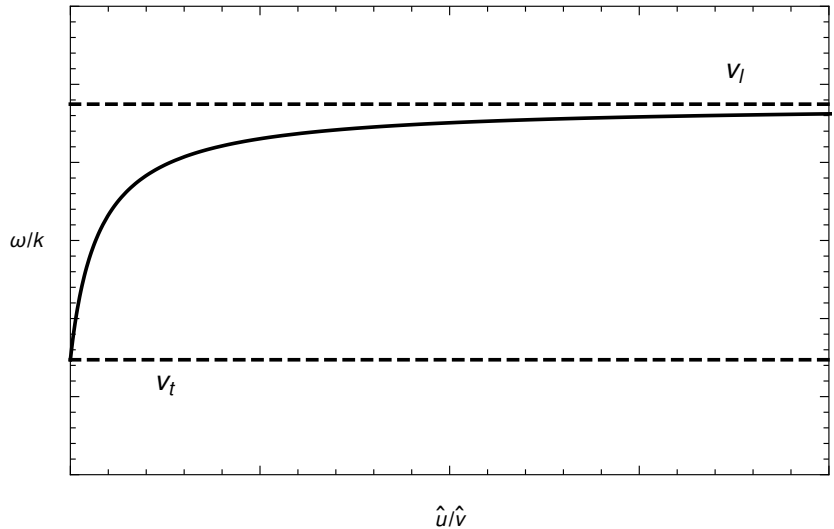}
  \caption{Visualisation of the dispersion relation (\ref{eq:dispersion}) as a function of the ratio of amplitudes $\hat{u}/\hat{v}$. The dashed lines indicate the limiting cases when $\hat{u}/\hat{v}\rightarrow 0$ (bottom) and $\hat{u}/\hat{v}\rightarrow \infty$ (top), respectively.}
  \label{fig:dispersion}
\end{figure}

Along the lines of \cite{boehmer2011rota,Liu20121907,boehmer2012gauge,boehmer2013rota,Boehmer2016158}, for instance, we can now study plane wave solutions of our more general model. For concreteness, let us consider a plane wave in the $x$-direction only. Then our ansatz for the displacements and microrotations is taken to be
\begin{align}
  \begin{pmatrix} u \\ v \\ \phi \end{pmatrix} = 
  \begin{pmatrix} \hat{u} \\ \hat{v} \\ \hat{\phi} \end{pmatrix} e^{ikx- i\omega t}\,,
  \label{eq:wave1}
\end{align}
where $k$ is the wave number and  $\omega$ is the angular frequency. The quantities $\hat{u},\hat{v},\hat{\phi}$ denote the corresponding amplitudes. Substitution of (\ref{eq:wave1}) into the equations of motion by replacing this form in (\ref{eom1})--(\ref{eom3}) and using the above renaming of constants $2d_1=\gamma$, $\mu^\ast=-\mu_c^\ast=A$, $\lambda^\ast=-2A$, $m_1/2+m_2=-A$, we can rewrite the equations in the following form
\begin{align}
  \begin{pmatrix}
    k^2 (\lambda +2 \mu )-\rho  \omega ^2 & -A k^2 & 2 i A k \\
    -A k^2 & k^2 (\mu_c +\mu )-\rho  \omega ^2 & -2 i k \mu_c \\
    -2 i A k & 2 i k \mu_c  & (\gamma  k^2+4 \mu_c + 4A) -\varrho_{\rm rot} \omega ^2 \\
  \end{pmatrix}
  \begin{pmatrix}
    \hat{u} \\ \hat{v} \\ \hat{\phi}
  \end{pmatrix} = 0\,.
\end{align}
As one would expect, the above equation is very similar to the one reported in \cite{Liu20121907} with the difference of the additional term $4A$ in the final component of the matrix. It is straightforward to show that the ratio of the amplitudes of the displacements is given by
\begin{align}
  \frac{\hat{u}}{\hat{v}} = 
  \frac{A \left(k^2 \mu - \rho \omega ^2\right)}
       {A^2 k^2 - \mu_c \left(k^2 (\lambda +2 \mu ) - \rho \omega ^2\right)} \,.
\end{align}
Next, let us find the dispersion relation using this ratio which gives
\begin{align}
  v = \frac{\omega}{k} = 
  \sqrt{\frac{\frac{\hat{u}}{\hat{v}}(\mu_c(\lambda+2\mu)-A^2)+A\mu}{\rho(\mu_c\frac{\hat{u}}{\hat{v}}+A)}}\,.
  \label{eq:dispersion}
\end{align}
The two interesting limits of this relation are when the ratio of the amplitudes is either very small or very large. In these cases we find
\begin{align}
  v_t := v_{\frac{\hat{u}}{\hat{v}} \rightarrow 0} =
  \sqrt{\frac{\mu}{\rho}}\,,
\end{align}
which is the well-known speed for the transversal elastic wave. Likewise,
\begin{align}
  v_l := v_{\frac{\hat{u}}{\hat{v}} \rightarrow \infty} =
  \sqrt{\frac{\lambda+2\mu}{\rho}-\frac{A^2}{\rho\mu_c}}\,,
\end{align}
so that we note that the longitudinal wave's speed is decreased by the chiral term $A$. The dispersion relation is shown in Fig.~\ref{fig:dispersion}. For this wave speed to be positive requires the constant $A$ to satisfy the inequality
\begin{align}
  A^2 > \mu_c (\lambda + 2\mu) \,. 
\end{align}

Finally, we can consider the specific case where we only have displacements for the horizontal direction $u$, so that $v=0$. In this case we use the ansatz $u=\hat{u}\cos(k x - \omega t + \delta_u)$ and $\phi = \hat{\phi}\cos(k x - \omega t + \delta_\phi)$. Provided the phase difference between the two waves is $\pi/2$, this means $\delta_\phi - \delta_u = \pi/2$, one can find a special solution to the linearised equations given by
\begin{align}
  \hat{u} = -\frac{2\mu_c}{A k}\hat{\phi}\,,\qquad
  \rho = \frac{k^2 \left(\mu_c(\lambda + 2\mu) - A^2\right)}{\mu_c\, \omega ^2}\,, \qquad
  \varrho_{\rm rot} = \frac{\gamma k^2 + 4 A}{\omega ^2}\,.
\end{align}
This is an interesting result as solutions of this type imply conditions that have to be satisfied regarding the parameters of the theory. There does not exist a non-trivial solution when $\mu_c=0$, in this case one finds $u=v=\phi=0$. This is consistent with theoretical considerations since the choice $\mu_c=0$ is not admitted in the linearised context. 

\subsection{Other possible applications}

Our intrinsic two-dimensional model might be particularly suited to describe materials like graphene and carbon nanotubes. These materials have unusual properties which make them valuable for applications in nanotechnology. The mathematical description of electrons propagating in graphene, for example, requires the use of the two-dimensional massless Dirac equation. The microstructure of Cosserat elasticity leans itself naturally to consider applications in this area and some research along those lines is ongoing. In~\cite{Caillerie2006} a homogenization was applied to model a graphene sheet, while applications of Cosserat elasticity were considered in~\cite{DeCicco2013} and also~\cite{doi:10.1063/1.4951692}. Elastic properties of graphene were studied in~\cite{PhysRevB.82.235414}. Carbon nanotubes, on the other hand, were studied in the context of Cosserat elasticity in~\cite{Selmi:2014}. It will be interesting to apply our approach to some of these models in the future.

\section{Conclusions}

The primary motivation of this paper was the formulation of an intrinsic two-dimensional, geometrically non-linear Cosserat theory of elasticity that could be used to model, amongst other features, chiral lattices. Our construction circumvents the problems one faces when reducing a three-dimensional isotropic and chiral theory to the plane. It is well known that the resulting two-dimensional theory is no longer chiral.

Our model is intrinsically two-dimensional and does not refer to a three-dimensional model for its construction. Chirality is introduced into this model by defining a new rotated deformation gradient $F^{\ast}$, see eq.~(\ref{fstar}), which is constructed from the deformation vector rotated by $\pi/2$ in the counter-clockwise direction. This deformation gradient is then used to define a new elastic energy and also new interaction terms between the deformation gradient $F$ and the new quantity $F^{\ast}$. This approach is different to other approaches as we are directly modelling the planar chiral material. 

Due to our intrinsic approach our model displays a certain flexibility in the sense that it depends on up to 10 different elastic constants and the characteristic length $L_c$. These are the two-dimensional elastic (Lam\'e) constants $\mu, \lambda$, the Cosserat couple modulus $\mu_c$, and the interaction term $\chi$. Moreover, we have the corresponding constants $\mu^{\ast}, \lambda^{\ast}, \mu^{\ast}_c$ and finally the three constants $m_1,m_2,m_3$. This allows us to formulate previously studied models using our intrinsic approach. It would also be interesting to apply our approach to other two-dimensional materials like graphene. 

Finally, let us draw the attention back to the question of how to obtain a chiral two-dimensional model. We have seen that using a quadratic ansatz in the three-dimensional Cosserat model necessarily cannot give a two-dimensional chiral model, therefore our effort. However, it is not clear whether this problematic issue can be avoided by using some higher order (non-quadratic) model in the first place. If this would be possible, the chiral effect would be absent under linearisation. Therefore, more experimental evidence is needed to decide whether chirality can and should appear already as a discernible effect in an infinitesimal neighbourhood of the identity. 

\section*{Acknowledgement}
SB is supported by the Comisi{\'o}n Nacional de Investigaci{\'o}n Cient{\'{\i}}fica y Tecnol{\'o}gica (Becas Chile Grant No.~72150066).

\appendix

\section{Chirality and the matrix Curl}
\label{app:chiral2}

Let us define the new deformation vector $\varphi^\#$ by
\begin{align}
  \varphi^\#(x,y,z) := \varphi(-x,-y,-z) \,.
\end{align}
This means we evaluate the function $\varphi$ at the inverted coordinate $(-x,-y,-z)$. However, we will work in the original Cartesian coordinate system and will not make the coordinate transformation of the previous appendix. Then, one verifies that
\begin{align}
  F^\# = \nabla \varphi^\# = - \nabla \varphi(-x,-y,-z) = -F(-x,-y,-z) \,,
\end{align}
because of the chain rule. Under this `inversion' the deformation gradient picks up a sign, making it non-invariant. On the other hand, $F^T F$ is invariant.

Next, let us define the new orthogonal matrix $R^\#$ which we define by
\begin{align}
  R^\# := -R(-x,-y,-z) \,,
\end{align}
where an additional minus sign was included in the definition. Then one immediately finds $(R^\#)^T = -R(-x,-y,-z)^T = -(R^T)(-x,-y,-z)$ and
\begin{align}
  \Curl R^\# =
  \begin{pmatrix}
    \partial_y R^\#_{xz} - \partial_z R^\#_{xy} &
    \partial_z R^\#_{xx} - \partial_x R^\#_{xz} &
    \partial_x R^\#_{xy} - \partial_y R^\#_{xx} \\
    \partial_y R^\#_{yz} - \partial_z R^\#_{yy} &
    \partial_z R^\#_{yx} - \partial_x R^\#_{yz} &
    \partial_x R^\#_{yy} - \partial_y R^\#_{yx} \\
    \partial_y R^\#_{zz} - \partial_z R^\#_{zy} &
    \partial_z R^\#_{zx} - \partial_x R^\#_{zz} &
    \partial_x R^\#_{zy} - \partial_y R^\#_{zx}
  \end{pmatrix} \,.
\end{align}
Each single derivative will yield a minus sign in virtue of the chain rule. For instance,
\begin{align}
  \partial_y R^\# := - \partial_y R(-x,-y,-z) = 
  \partial_{-y} R(-x,-y,-z) = (\partial_y R)(-x,-y,-z)\,.
\end{align}
Hence, applying this through all terms in the matrix Curl of $R^\#$, one arrives at
\begin{align}
  \Curl R^\# = (\Curl R)(-x,-y,-z) \,.
\end{align}
Finally, one can study the curvature measure $(R^\#)^T \Curl R^\#$ which satisfies
\begin{align}
  (R^\#)^T \Curl R^\# = -(R^T)(-x,-y,-z) (\Curl R)(-x,-y,-z) = 
  -(R^T \Curl R)(-x,-y,-z)\,.
\end{align}
This result shows the desired chiral properties. Consequently, we showed that the term $\langle F^T F, R^T \Curl R \rangle$ is chiral.

\section{Variations and equations of motion}
\label{app:vari}

Each of the four parts of our energy functional (\ref{fullenergy}) and (\ref{chiral}) are considered separately. The various matrix derivatives that occur in this calculation are computed using standard formulae, all of which can be found nicely presented in the {\it The Matrix Cookbook}~\cite{Mcookbook}.

\subsection{Elastic energy $V_{\rm elastic}$}

We can rewrite the elastic energy functional as
\begin{align}
  V_{\rm elastic} = {}&{}
  \mu\, \|\sym \overline{R}^{T}F-\id\|^{2} +
  \frac{\lambda}{2} \left(\tr(\sym(\overline{R}^{T}F)-\id)\right)^{2}
  \nonumber \\ = {}&{} 2\mu - 2\mu \tr(F\overline{R}^{T}) + 
  \frac{\mu}{2} \left(\tr(\overline{R}^{T}F\overline{R}^{T}F) + \tr(FF^{T})\right) 
  \nonumber \\ &{}+
  2\lambda -2\lambda \tr(\overline{R}^{T}F)+ 
  \frac{\lambda}{2}\left(\tr(\overline{R}^{T}F)\right)^{2} \,.
  \label{el111}
\end{align}
Now we can compute the variations of this energy functional
\begin{align}
  \delta V_{\rm elastic}(F,\overline{R}) = {}&{} 
  \frac{\mu}{2}\Big(\frac{d}{d\overline{R}}\Big[\tr(\overline{R}^{T}F\overline{R}^{T}F)\Big]:
  \delta \overline{R} + 
  \frac{d}{dF}\Big[\tr(\overline{R}^{T}F\overline{R}^{T}F)+\frac{d}{dF}(\tr(FF^{T}))\Big]:
  \delta F\Big) 
  \nonumber\\ &{}- 
  2(\mu+\lambda)\Big(\frac{d}{d\overline{R}}(\tr(\overline{R}^{T}F)):\delta \overline{R}+\frac{d}{dF}(\tr(\overline{R}^{T}F)):\delta F\Big)
  \nonumber\\ &{}+
  \frac{\lambda}{2}\Big(\frac{d}{d\overline{R}}[\tr(\overline{R}^{T}F)]^{2}:\delta \overline{R}+
  \frac{d}{dF}[\tr(\overline{R}^{T}F)]^{2}:\delta F\Big) \,.
  \label{el333}
\end{align}
Now, computing the various matrix derivatives yields
\begin{align}
  \delta V_{\rm elastic}(F,\overline{R}) = {}&{} 
  \bigl\langle
    \mu(\overline{R}F^{T}\overline{R}+F) -
    2(\mu+\lambda)\overline{R} + 
    \lambda \,\tr(\overline{R}^{T}F)\overline{R}
  ,\delta F \bigr\rangle
  \nonumber \\ &{} +
  \bigl\langle
    \mu F\overline{R}^{T}F -
    2(\mu+\lambda)F + 
    \lambda \, \tr(\overline{R}^{T}F)F
  ,\delta \overline{R} \bigr\rangle \,.
  \label{deltaVelastic11}
\end{align}

If we want to study the dynamical problem, we will need to subtract the kinetic energy term
\begin{align}
  V_{\rm elastic,kinetic} &= \frac{\rho}{2} \|\dot{\varphi}\|^{2} \,.
\end{align}
Here $\rho$ is the density. If we vary this term we will obtain $\delta V_{\rm elastic,kinetic} = -\rho\, \ddot{\varphi}\, \delta\varphi$, where we neglected a boundary term when integrating by parts once. Next, using that $\nabla\varphi=\mathbbm{1}+\nabla u$ implies $\delta \varphi =\delta u$. Hence, the elastic kinetic term can finally be written as
\begin{align}
  \delta V_{\rm elastic,kinetic} = -\rho\,\ddot{u}\, \delta u \,.
\end{align}

\subsection{Curvature energy $V_{\rm curvature}$}

The curvature energy functional depends only on the rotation angle $\vartheta$ so that 
\begin{align}
  V_{\rm curvature}(\vartheta) = \mu L_c^2 \, \|\grad\vartheta\|^{2} \,.
  \label{transss}
\end{align}
Variations with respect to $\delta\vartheta$ are well known (we are dealing with a standard vector), and one finds
\begin{align}
  \delta V_{\rm curvature} = -2 \mu L_c^2 \divv(\grad\vartheta) \, \delta\vartheta =
  -2 \mu L_c^2\, \Delta \vartheta \, \delta\vartheta \,,
  \label{deltaVCUR2}
\end{align}
where, as before, we have neglected a boundary term. Recall that $\divv\grad\vartheta$ is the scalar Laplacian $\Delta \vartheta$. If we want to study the dynamical problem, we will need to subtract the kinetic energy term
\begin{align}
  V_{\rm curvature,kinetic} = \rho_{\rm rot} \|\dot{\vartheta}\|^{2} \,.
\end{align}
Variations with respect to $\delta\vartheta$ lead to
\begin{align}
  \delta V_{\rm curvature,kinetic} = -2\rho_{\rm rot} \ddot{\vartheta}\, \delta\vartheta \,,
\end{align}
up to a boundary term.

\subsection{Interaction energy $V_{\rm interaction}$}

The interaction term is
\begin{align}
  V_{\rm interaction}(F,\overline{R}) = 
  \mu L_c \chi \|\overline{R}^{T}\! \textrm{Curl}\overline{R}\| \tr(\overline{R}^{T}F) =
  \mu L_c \chi \|\grad\vartheta\| \tr(\overline{R}^{T}F) \,.
\end{align}
Computing the variations with respect to $\overline{R}$ and $F$ yields
\begin{align}
  \delta V_{\rm interaction}(F,\vartheta) = {}& 
  \mu L_c \chi \tr(\overline{R}^{T}F) \frac{\grad\vartheta}{\|\grad\vartheta\|}\delta (\grad\vartheta) 
  \nonumber\\ &+ 
  \mu L_c \chi \|\grad\vartheta\| \Big(\frac{d}{dF}[\tr(\overline{R}^{T}F)]:\delta F +
  \frac{d}{d\overline{R}}[\tr(\overline{R}F^{T})]:\delta \overline{R}\Big) 
  \nonumber \\ = {}& -
  \mu L_c \chi \frac{\grad\vartheta}{\|\grad\vartheta\|}\grad\Big[\tr(\overline{R}^{T}F)\Big]\delta\vartheta -
  \mu L_c \chi \,\tr(\overline{R}^{T}F)\divv\Big[\frac{\grad\vartheta}{\|\grad\vartheta\|}\Big]\delta\vartheta
  \nonumber\\ &+
  \mu L_c \chi\, \bigl\langle \|(\grad\vartheta)\|F, \delta \overline{R}\bigr\rangle +
  \mu L_c \chi\, \big\langle \|(\grad\vartheta)\|\overline{R},\delta F\bigr\rangle \,,
  \label{deltaVinteraction}
\end{align}
which is probably the most complicated of all terms.

\subsection{Cosserat couplings $V_{\rm coup}$}

We have the energy functional of the coupling as follows
\begin{align}
  V_{\rm coupling}(F,\overline{R}) &= \mu_{c}\, \|\overline{R}^{T}\polar(F)-\id\|^{2} = 
  4\mu_{c} - 2\mu_{c}\tr(\overline{R}^{T} \polar(F)) \,.
  \label{Vcoup2}
\end{align}
The variations of $\polar(F)$ with respect to $F$ are somewhat non-standard, however, the result is well-known and has been reported for instance in~\cite{ChenWheeler1993,Rosati1999}. In particular, in two dimensions, the result is straightforward to verify directly. We briefly recall that the polar part of $F$ can be written as $\polar(F)=F(F^T F)^{-1/2}$. Then
\begin{align}
  \delta V_{\rm coupling}(F,\overline{R}) = - 
  2\mu_{c}\polar(F):\delta \overline{R} - 
  2\mu_{c}\Big[\frac{d}{dF}\Big(\tr (\overline{R}^{T}\polar(F))\Big)\Big]:\delta F \,.
\end{align}
Using the chain rule for matrix differentiation we find
\begin{align}
 \frac{d}{dF_{ml}} \Big(\tr(\overline{R}^{T}\polar(F))\Big) &=
 \tr
 \left[
   \left(\frac{d}{d\polar(F)}\tr(\overline{R}^{T}\polar(F))\right)^{T}\frac{d\polar(F)}{dF_{ml}}
 \right]
 \nonumber \\ &=
 \tr
 \left[
   \overline{R}^T \frac{d\polar(F)}{dF_{ml}}
 \right] \,.
\end{align}

Following~\cite{ChenWheeler1993,Rosati1999} we have
\begin{eqnarray}
  \frac{d\polar(F)}{dF} = 
  \frac{1}{\tr(U)}(\mathds{I} - \polar(F)\, \hat{\boxtimes}\, \polar(F)^{T}) \,,
\end{eqnarray}
where we followed the notation used in~\cite{Rosati1999}. Recall that $U$ is the positive definite symmetric part of the polar decomposition $F=\polar(F)\,U$. The symbol $\hat{\boxtimes}$ denotes the operation $(A \hat{\boxtimes} B)_{ijkl} = A_{il} B_{jk}$ for any two matrices $A$ and $B$. In addition we use the notation $\mathds{I}=\id \hat{\boxtimes} \id$. Putting this together yields
\begin{align}
  \tr
  \left[
    \overline{R}^T \frac{d\polar(F)}{dF_{ml}}
    \right] = 
  \frac{1}{\tr(U)} \left[\overline{R}-R\overline{R}^{T}R\right] \,.
\end{align}
Finally, we will obtain the variations $\delta V_{\rm coupling}$ which are given by 
\begin{align}
  \delta V_{\rm coupling}(F,\overline{R}) = -
  2\mu_{c}\, \bigl\langle R, \delta \overline{R}\bigr\rangle - 
  \frac{2\mu_{c}}{\tr(U)}
  \big\langle \overline{R}-R\overline{R}^{T}R ,\delta F \big\rangle \,.
  \label{deltaVcoupling2}
\end{align}

Now we will focus on the second coupling term (\ref{Vcoupagain2}) that can be rewritten as follows
\begin{align}
  V_{\rm coupling(2)}(F,\overline{R})=
  3\mu_c + \frac{\mu_c}{2}
  \tr \Bigl[ F^{T}F - \overline{R}^{T}F \overline{R}^{T}F \Bigr] \,,
\end{align}	
which yields
\begin{align}
  \delta V_{\rm coupling(2)}(F,\overline{R}) =
  \mu_c \bigl\langle F-\overline{R}F^{T}\overline{R},\delta F\bigr\rangle -
  \mu_c \bigl\langle \overline{R}^{T}F,\delta \overline{R}\bigr\rangle \,.
\end{align}

\subsection{Chiral terms}

Let us begin with computing the term $\delta V_{\rm elastic}^{\ast}$. A direct calculation yields
\begin{align}
  \delta V^{\ast}_{\rm elastic} = &{} 
  -\Divv\Big[
    \mu^{\ast}(\overline{R}(F^{\ast})^{T}\overline{R}+F^{\ast}) -
    2(\mu^{\ast}+\lambda^{\ast})\overline{R} + 
    \lambda^{\ast} \,\tr(\overline{R}^{T}F^{\ast})\overline{R}
    + \mu_c^{\ast} (\overline{R}(F^{\ast})^{T}\overline{R}-F^{\ast})
    \Big]:\epsilon\ \delta u 
  \nonumber \\ &+
  \Big[
    \mu^{\ast} F^{\ast}\overline{R}^{T}F^{\ast} -
    2(\mu^{\ast}+\lambda^{\ast})F^{\ast} + 
    \lambda^{\ast} \, \tr(\overline{R}^{T}F^{\ast})F^{\ast} +
    \mu_c^{\ast} F^{\ast}\overline{R}^{T}F^{\ast}
    \Big]:(-\epsilon \overline{R})\ \delta \vartheta \,.
  \label{deltaVelasticstar}
\end{align}
Next, we are considering the mixing terms
\begin{align}
  \delta V_{\rm mixing} = {}&{}
  \frac{1}{2}m_{1}\Big(\delta\Big[\tr(\overline{R}^{T}F^{\ast}\overline{R}^{T}F)\Big] +
  \delta\Big[\tr(F(F^{\ast})^{T})\Big]-
  2\delta\Big[\tr(\overline{R}^{T}F^{\ast})+\tr(\overline{R}^{T}F)\Big]\Big)
  \nonumber\\ &+
  m_{2}\Big(\delta\Big[\tr(\overline{R}^{T}F^{\ast})\tr(\overline{R}^{T}F)\Big]-\delta\Big[\tr(\overline{R}^{T}F^{\ast})+\tr(\overline{R}^{T}F)\Big]\Big)
  \nonumber\\ &+
  \frac{1}{2}m_{3}\Big(\delta\Big[\tr(F(F^{\ast})^{T})\Big]-\delta\Big[\tr(\overline{R}^{T}F^{\ast}\overline{R}^{T}F)\Big]\Big)\,.
  \label{Vmixing}
\end{align}
These variations are slightly more involved than the previous ones, so some additional details are given in the following. The first three respective terms are given by
\begin{align}
  \delta\Big[\tr(\overline{R}^{T}F^{\ast}\overline{R}^{T}F)\Big] &=
  (F\overline{R}^{T}\epsilon^{T}F+\epsilon^{T}F\overline{R}F):\delta \overline{R} +
  (\epsilon^{T}\overline{R}F^{T}\overline{R}+\overline{R}F^{T}\epsilon^{T}\overline{R}):\delta F\,,
  \\
  \delta\Big[\tr(F(F^{\ast})^{T})\Big] &=
  F:\delta F^{\ast}+F^{\ast}:\delta F =\Big[(F:\epsilon)+F^{\ast}\Big]:\delta F\,,
  \\
  \delta\Big[\tr(\overline{R}^{T}F^{\ast})+\tr(\overline{R}^{T}F)\Big] &=
  (\overline{R}:\epsilon+\overline{R}):\delta F+(F+F^{\ast}):\delta \overline{R} \,.
\end{align}
The fourth and final term are given by
\begin{multline}
  \delta\Big[\tr(\overline{R}^{T}F^{\ast})\tr(\overline{R}^{T}F)\Big] =
  (\tr(\overline{R}^{T}F^{\ast})F+\tr(\overline{R}^{T}F)F^{\ast}):\delta R\\ +
  (\tr(\overline{R}^{T}F^{\ast})R+\tr(\overline{R}^{T}F)(R:\epsilon)):\delta F\,.
\end{multline}
Therefore, the complete variations are given by
\begin{align}
  \delta V_{\rm mixing} = {}&{}
  \Bigl[
    \frac{1}{2}m_{1}
    \Big(F\overline{R}^{T}\epsilon^{T}F+\epsilon^{T}F\overline{R}F-2(F+F^{\ast})\Big)
    \nonumber \\ &+
    m_{2}\Big(\tr(\overline{R}^{T}F^{\ast})F+\tr(\overline{R}^{T}F)F^{\ast}-(F+F^{\ast})\Big)
    \nonumber\\ &-
    \frac{1}{2}m_{3}\Big(F\overline{R}^{T}\epsilon^{T}F+\epsilon^{T}F\overline{R}F\Big)\Bigr]:\delta \overline{R}
  \nonumber \\ &+
  \Bigl[
    \frac{1}{2}m_{1}\Big(\epsilon^{T}\overline{R}F^{T}\overline{R}+\overline{R}(F^{\ast})^{T}\overline{R}+F:\epsilon+F^{\ast}-2(\overline{R}:\epsilon+\overline{R})\Big)
    \nonumber \\ &+
    m_{2}\Big(\tr(\overline{R}^{T}F^{\ast})R+\tr(\overline{R}^{T}F)(R:\epsilon)-(\overline{R}:\epsilon+\overline{R})\Big)
    \nonumber \\ &+
    \frac{1}{2}m_{3}\Big(F:\epsilon+F^{\ast} -(F\overline{R}^{T}\epsilon^{T}F+\epsilon^{T}F\overline{R}F)\Big)
    \Big]:\delta F\,.
\end{align}

\end{document}